\documentstyle[aps,prl,epsfig,twocolumn]{revtex}
\title{Proposal for high-precision Atomic Parity Violation measurements\\
using amplification of the asymmetry by stimulated emission\\
in a transverse $\vec E$ and $\vec B$ field pump-probe experiment }
 \author{J. Gu\'ena, M. Lintz and M. A. Bouchiat $^*$  }
\address{ Laboratoire Kastler Brossel $^a$ et F\'ed\'eration de Recherche $^b$ \\D\'epartement de Physique de l'Ecole
Normale Sup\'erieure,\\ 24 Rue Lhomond, F-75231  Paris  Cedex 05, France
\\
 (\today)} 
\author{\parbox{410pt}{ \vglue 0.3cm \footnotesize\sf
 Amplification by stimulated emission of radiation provides an intriguing means for increasing the sensitivity of Atomic Parity
Violation (APV) measurements in a pump-probe configuration well adapted to the $6S-7S$ cesium transition. It takes
advantage of the large number of atoms excited along the path of the pump beam. In the longitudinal $E$- field configuration
currently exploited in an ongoing APV measurement, this number is limited only by the total voltage sustainable by the Cs
vapor. In order to overcome this limit, we consider, both theoretically and experimentally, the possibility of performing the
measurements in {\it a transverse $E$-field} configuration requiring a much lower voltage. We discuss the necessarily different
nature of the observable and the magnetoelectric optical effects entering into play. They condition modifications of the experimental
configuration with, in particular, the application of a transverse magnetic field. We suggest the possibility of rotating the 
transverse direction of the fields so as to suppress systematic effects. With {\it a long interaction length}, a  
precision reaching $0.1\%$ in a quantum noise limited measurement can be expected, now limited only by the necessity of operating
below the threshold of spontaneous superradiant emission of the excited medium. If we approached this limit, however, we could
greatly amplify the asymmetry using {\it triggered} superradiance.  PACS numbers: 32.80.Ys, 11.30.Er, 33.55.b 
 }\vspace{-1.5cm}}
   
\begin{document}
\maketitle
\section {Introduction}
\subsection{Motivation for new independent measurements} 
The main goal of Atomic Parity Violation (APV) experiments is a precise determination of $Q_W$, the so-called weak charge of
the nucleus  which, as far as the electron-nucleus weak interaction is concerned, plays the same role as the electric
charge for the Coulomb interaction \cite{bou97}. The value of $Q_W(Cs)$ is quoted as a significant constraint on ``New Physics''
in the Particle Data Book \cite{PDB02}. Atomic calculations in cesium have reduced their uncertainty below
$1\%$ \cite{dzu02} and many cross checks have reinforced confidence in the results at this level \cite{blu90,der01,mil01}.
On the experimental side a single group \cite{woo97,ben99} has met the challenge which consists in achieving a calibrated
\cite{bou881} measurement  of the APV $6S-7S$ transition amplitude $E_1^{pv}$ at a level of precision better than $1\%$.
However, the  measurement of any fundamental quantity has to pass the test of the verification by a completely independent
method. Moreover, it has been reemphasized recently \cite{joh03,CB91} that their reported value for the small  nuclear
spin-dependent contribution is difficult to explain theoretically.
 
APV experiments are notoriously difficult from the standpoints of both low statistics and the possibility of 
numerous systematic effects. Solutions which increase the Signal to Noise Ratio (SNR) often give rise to undesirable spurious
effects. In our first Cs APV Stark interference experiment \cite{bou82}, the detection method was selective but
inefficient.  The same was true for APV  measurements in thallium \cite{com79}. Conversely, in
 the atomic beam experiments on Cs \cite{woo97}, the SNR is huge by comparison, thanks to  the high light intensity 
obtained in the power build-up cavity. However, this is at the price of non-linear effects 
which complicate interpretation of the data.

\subsection{Interesting features of a pump-probe experiment  }
More recently, another method \cite{bou85} has provided its first APV measurements \cite{bou02}, and highlighted a
mechanism looking attractive as regards its potential gain of sensitivity, namely amplification by stimulated emission of
radiation  
\cite{bou96}. It relies on a pump-probe scheme in which the excited $7S$ state is populated by an intense, 
pulsed excitation beam, with a linear polarization $\hat \epsilon_{ex}$ and a wave vector $\hat k$ parallel to the applied electric
field $\vec E$. The atoms are then probed by a second, collinear pulsed laser beam which is tuned to resonance for the
$7S\rightarrow 6P_{3/2}$ transition. The probe beam is amplified by stimulated emission. At the same time its linear
polarization $\hat
\epsilon_{pr}$, undergoes a small change of direction or pseudo-rotation, due to a linear dichroism arising from the presence of
the pseudoscalar term   
$ (\hat \epsilon_{ex} \cdot \hat \epsilon_{pr})( \hat \epsilon_{ex} \wedge \hat \epsilon_{pr}\cdot \vec E)$ in the gain and
leading to a tilt $\theta^{pv}$ of its eigen-axes with respect to the planes of symmetry of the experiment \cite{bou02,gue98}.
This tilt is given by the ratio of the PV  E$_1$ amplitude to the vector part of the Stark induced amplitude~\cite{bou97}  
  $\theta^{pv} = - \rm{Im E}_1^{pv}/\beta {\it E} $, about 
1 $\mu$rad in current experimental conditions.  This is the important parameter to be measured. We note that the
pseudo-rotation of $\hat \epsilon_{pr}$  is, as $\theta^{pv}$, odd under 
$\vec E$ reversal, which provides a powerful tool for discriminating against systematic effects. We measure the rotation with a
high sensitivity, two-channel polarimeter operating in balanced mode
\cite{gue97}, in which it gives rise to an imbalance $D_{amp}=\frac {S_1-S_2}{S_1 + S_2}$ between the signals, $S_1, S_2$
recorded in the two channels
 detecting the {\it amplified} probe beam. To reject the
contribution due to imperfections of the probe polarization, the imbalance is also measured with a subsequent {\it reference} probe
pulse and subtracted \cite{noteA} to yield, at each excitation pulse, the imbalance of atomic origin, that is the left-right
asymmetry:
$A_{LR}
\equiv D_{at} = D_{amp} - D_{ref}$. Using real-time calibration by a known tilt of the optical axes of parity-conserving origin,
we deduce the value of $\theta^{pv}$.    
   
 We have already demonstrated that this method has a good detection efficiency (all photons arising from stimulated emission by
the probe beam reach the polarimeter) while preserving the selectivity of the fluorescence detection approach. In addition, this
scheme allows us to exploit rotational symmetry about the pump probe common beam axis
  to reduce systematic effects
\cite{bou03}. The limit of sensitivity achievable by our present technique is the important issue at stake, discussed
in this paper. The SNR is definitely better than that obtained using polarized fluorescence detection and it is
expected to bring us to within reach of our 1$\%$ precision objective.  However, one should not conclude that this represents  the
ultimate  sensitivity limit obtainable with the stimulated-emission method of detection. Indeed the key 
point here is that we use an enhancement mechanism which considerably increases both the
probe intensity {\it and} its left-right asymmetry with the optical density at the probe wavelength. Since we first validated the
method \cite{bou02},  we have checked that this mechanism does really increase the SNR. It has indeed contributed significantly to
 the increase by a factor of 3.3 that we have recently obtained. This raises the question of just how much more one might
expect to increase the sensitivity in this way. 
 
\subsection{Practical limit to the amplification with a longitudinal  $\vec E$ field} 
The optical density of the Cs vapor at the probe wavelength ${\cal{A}} = \ln {(n_{out}/n_{in})}$, deduced from the number
of incoming ($n_{in}$) and outgoing ($n_{out}$) probe photons, is proportional to the  number of $7S$ excited atoms in the
interaction region
$N$, itself proportional to the excitation beam energy, to the square of the electric field strength $E^2$ and to the length of the
cell.   At the quantum limit, the inverse of the noise equivalent angle $NEQA$ per pulse, hence the measurement sensitivity,
involves the square root of the number of probe photons per pulse reaching the detector, $n_{in}
\exp {({\cal{A}})}$, times the asymmetry amplification factor \cite{bou96}, $d A_{LR}/d \theta$. 
This result \cite{cha98} relies on two reasonable assumptions that we wish to make explicit here. First, the
polarimeter operates in balanced mode. Second, the number of collected photons spontaneously emitted in the mode of
the probe beam is very small as compared to the number of stimulated photons ($\simeq 10^{-6}$ in typical conditions), and
remains small at moderate amplifications, so that the associated noise can be neglected \cite{noteF}.  

In the limit $ {\cal{A}} \leq 1 $, the asymmetry amplification
 can be expressed in an identical way for both the ``para''  ($\hat \epsilon_{ex} \parallel \hat
\epsilon_{pr}$), and  ``ortho'' ($\hat \epsilon_{ex} \perp \hat
\epsilon_{pr}$), configurations:
\begin{eqnarray}   
  \hspace{-40mm}d A_{LR}/d \theta =  2\eta_{av}{\cal{A}}_{av} &\,,& \\
{\rm with } \hspace {3 mm} {\cal{A}}_{av} =
({\cal{A}}_{\parallel}+{\cal{A}}_{\perp})/2 \hspace{3mm} &{ \rm and}& \hspace{3mm}\eta_{av}= \frac {\alpha_{\parallel} -
\alpha_{\perp}}{\alpha_{\parallel} + \alpha_{\perp}}\, , \nonumber 
\end{eqnarray} 
 where ${\cal{A}}_{\parallel}$ and ${\cal{A}}_{\perp}$ denote the optical
densities and
$\alpha_{\parallel}$ and
$\alpha_{\perp}$ the amplification coefficients for the field per unit length, in the two configurations
\cite{bou96}. For the favorable $6S_{F=3} \rightarrow 7S_{F=4}
\rightarrow 6P_{3/2,F=4}$ transition $ 2\eta_{av}= 22/23 $, which henceforth we approximate by 1. Thus
we can write:      
\begin{equation}
 (NEQA)^{-1} = \sqrt {n_{in}} \exp {({\cal{A}}_{av}/2)} \times {\cal{A}}_{av}\, , 
\end{equation}
which obviously corresponds to a rapid increase of the quantum noise limited SNR
$ =\theta^{pv} (NEQA)^{-1}$/pulse as a function of
${\cal{A}}$. 

 In the longitudinal field configuration, where $\vec E_{\parallel}$ lies along the common beam direction  $\hat k$, the only 
parameter left in our experiment for increasing ${\cal {A}}$ is the excitation beam intensity. 
 To increase the applied electric field we need to increase the applied potentials but at the large Cs vapor densities at which
we operate (a few times $ 10^{14 }$/cm${^3}$), too large a voltage would present the risk of a discharge inside the vapor.  With
$\vec E  \parallel  \hat  k$, increasing the length also requires a prohibitive  increase of the applied voltage. By contrast, this kind
of  limitation is obviously absent if the experiment is performed not in the longitudinal but in the {\it transverse} field
configuration. We also note that in a transverse field $\vec E_{\perp}$, the left-right PV asymmetry receives a contribution not
only from the vector Stark induced amplitude  $i\beta \vec E_{\perp } \wedge \vec \sigma
\cdot \hat \epsilon_{ex}$ (where $\vec \sigma$ denotes the electron spin operator) but also from the scalar one $\alpha \vec
E_{\perp} \cdot \hat \epsilon_{ex}$.  For the $6S \rightarrow 7S$ Cs transition \cite{bou97}, 
$\vert \alpha \vert 
\simeq 10 \,\vert \beta\vert $. Therefore, at smaller field magnitudes $E_{\perp} $, typically close to $E_{\parallel}/10
$, and with a longer interaction length, we can expect an overall left-right asymmetry
significantly larger than in the longitudinal field configuration. However, the physical quantity which bears the PV signature in 
a transverse E-field is different. We examine it in detail in the next section.

\section{A pump-probe APV experiment\\  in a transverse electric field}
\subsection{The need for a magnetic field}
While in a longitudinal $\vec E$ field the observables are the different contributions to the $7S$-excited alignment
\cite{bou03}, detected through the linear dichroism or birefringence to which they give rise, in the transverse $\vec E$ field
geometry the observables are the contributions to the 
$7S$ orientation giving rise to circular dichroism or optical rotation. However, only the orientation component $\vec
P\cdot \hat k$ collinear to the beam contributes to the probe polarization signal. We show below that it is possible to realize
conditions such that, at least in principle, only the orientation of parity violating origin participates to the signal. 

 There are several contributions to the orientation created in the
$7S$ state. The interference between the Stark-induced electric-dipole transition amplitudes of scalar and vector origins
gives rise to a longitudinal component  $\vec P^{(2)} = -\frac{5}{6} \frac{\beta}{\alpha} \xi_{ex}
\hat k$,  where $\xi_{ex}$ denotes the photon helicity. In addition, the interference of the scalar Stark-induced amplitude
with the magnetic-dipole amplitude $M_1$ on the one hand and the parity violating electric-dipole amplitude $ \rm{Im
E}_1^{pv}$ on the other, gives rise to the transverse component:
\begin{equation}
\vec P_{\perp} = \frac{5}{6}\frac{ M_1 + \xi_{ex} {\rm Im\,E}_1^{pv}}{
\alpha E_{\perp}}  ( \hat k \wedge \hat E_{\perp}) \equiv \vec P^{(1)} + \vec P^{pv},  
 \end{equation}
 after excitation of the $6S_{F=4}  \rightarrow 7S_{F=4} $ line.
We have split $\vec P_{\perp}$ into the PV contribution $\vec P^{pv}$ and   
the $M_1$ contribution $\vec P^{(1)}$ which is 2$\times 10^4$ larger, but which cancels out if the
atoms are excited by two counterpropagating beams of equal intensity \cite{noteE} (see Fig.1).
\vspace{10mm}

\begin{figure}
\centerline{\epsfxsize=42mm \rotatebox{90}{\epsfbox{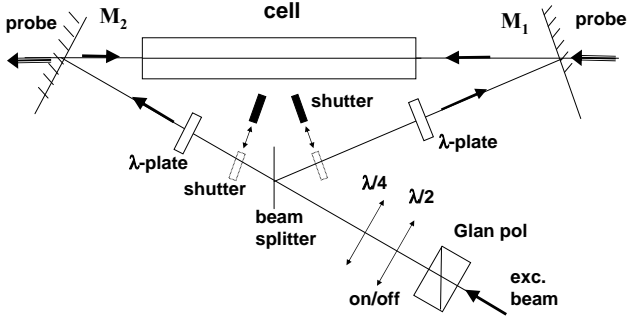}}}
\vspace{5mm}
\caption{\footnotesize Schematic of the optical arrangement allowing for the cancellation of $\vec P^{(1)}$: two
counterpropagating excitation beams are made colinear to the probe beam by using dichroic mirrors ($M_1,\, M_2$); their
extinction is monitored by two independent shutters. The birefringences on the two paths are corrected by adjusting the axis 
orientation and the tilt of two $\lambda$-plates.  }
\end{figure} 

Since the $ \rm{Im\, E}_1^{pv}$ contribution we want to observe is perpendicular to $\hat k$ and to $\vec
E_{\perp}$, we apply a magnetic field $\vec B$ in the direction parallel to $ \vec E_{\perp}$. We also delay the probe pulse with
respect to the excitation pulse to give the transverse  orientation created sufficient time to precess around
$\vec B$ towards the direction of the beam $\hat k$. Simultaneously, the component $\vec P ^{(2)}$ precesses by the same
angle and becomes
 orthogonal to $\hat k$. If the instants of excitation and detection for all atoms coincided, we would expect the simple
time-dependences $ \sin({\omega_L} t)$ for $\vec P^{pv}$  and
$\cos({\omega_L} t)$ for $\vec P^{(1)}$. By numerically averaging over realistic time distributions we find  
the predicted time-dependence is still described by a sine and a cosine law to within a good approximation, if one replaces the
Larmor angular precession frequency  $\omega_L$ of one atom in the applied
$\vec B$ field by an effective angular frequency $\omega_{eff}$ whose numerical value depends on experimental
parameters such as excitation pulse duration and probe intensity.

\subsection{Dark-field detection}
During an exploratory stage of the experiment \cite{jah01a}, we have observed the $B$-field dependence of $\vec P^{(1)} \cdot
\hat k$ and
$\vec P^{(2)}\cdot \hat  k$ (Fig.2). We have found that the field magnitude required to suppress the $\vec P^{(2)}\cdot \hat k$
Stark-induced longitudinal contribution is 28~G for an excitation-probe pulse delay of 20 ns, {\it i.e.} sufficiently short to avoid
significant loss of excited atoms by spontaneous decay. Another important feature is that,  for that delay, the magnitude of
$\vec P^{(1)}$ observed with a single excitation beam is very close to its maximum value. Therefore, adjustment of both the delay
and the field magnitude combined with the use of two counter-propagating beams enables us to ensure dark-field detection of
$\vec P^{pv}$. We have verified that, over durations as short as 20 ns, the damping of the 7S polarization due to collisional
relaxation is completely negligible.
\begin{figure}
\centerline{\epsfxsize=70mm  \epsfbox{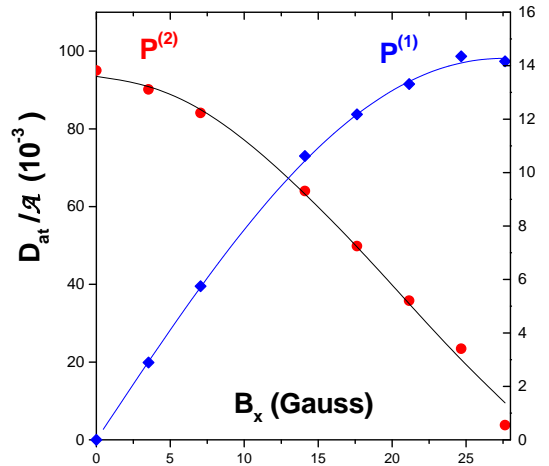}}
\vspace{8mm}
\caption{\footnotesize Larmor precession of $\vec P^{(1)}$ and $\vec P^{(2)}$ in a magnetic field $\vec B$ parallel to  $\vec
E_{\perp}$, observed in real time with a single excitation beam: the pump-probe delay is kept fixed at 20 ns and the
magnitude of the field is varied.
$\vec P^{(1)} \cdot \hat k$ (diamonds, right scale) and $\vec P^{(2)} \cdot \hat k$ (dots, left scale) are deduced from the
contributions to
$D_{at}/{\cal{A}}$ (for definitions see the text) respectively
$\xi_{ex}$-even,
$E_{\perp}$-odd, B-odd and  $\xi_{ex}$-odd, $E_{\perp}$-even, B-even. ($E_{\perp}$= 185 V/cm; exc. pulse 0.9 mJ;
resonant for $6S_{F=4} \rightarrow 7S_{F=4}$; beam diameter 1.8 mm;  probe pulse, $4\times 10^7$  linearly polarized
($\parallel  \vec B$) photons, resonant for $7S_{F=4}\rightarrow 6P_{3/2, F=5} $) \protect\cite{jah01a}.         }  
\end{figure}

The linearly polarized resonant probe beam passing through the oriented vapor acquires a small helicity.
For sensitive detection of this helicity we have converted our two-channel
polarimeter \cite{gue97} from a linear to a circular analyzer by inserting a quarter wave plate in front of the polarizing beam
splitter cube. The imbalance of atomic origin is still defined as $D_{at} = D_{amp} - D_{ref}$.   The tiny contribution
to $D_{at}$ which is odd under the  reversals of $\xi_{ex}$,
$\vec E$ and $\vec B$  provides us with the PV observable, $\vec P^{pv}\cdot \hat k$. An important step is to normalize 
it by another imbalance of well known atomic origin serving as a calibration. For this purpose, we can use the imbalances
occuring when one or the other of the two counterpropagating excitation  beams is blocked. They provide us with the
observable
$\vec P^{(1)}\cdot \hat k$. Thereby we can extract directly ${\rm Im\, E}_1^{pv} / M_1$ from the ratio of the
$E$-odd atomic imbalance measured with both excitation beams present, $D^{at} (\hat k_{ex}, -\hat k_{ex})$, to the
difference of the $E$-odd imbalances obtained with a single excitation beam present, 
$D^{at} (\hat k_{ex}) - D^{at}( -\hat k_{ex})$. It is interesting to note that this normalization procedure has the advantage
of eliminating the magnitude of the electric field. However, because of the asymmetry amplification process, for precise
measurements it is important to measure all the imbalances to be compared at equal optical densities. One simple way to
fulfill this condition  for measurements made alternatively with a single beam or two beams, might 
consist in increasing the applied potential for a single beam so as to double $E_{\perp}^2$, the extracted quantity
then being $\sqrt 2 \,  ({\rm Im\, E}_1^{pv} / M_1) $.       

 It can be recalled that a PV asymmetry of the same origin has already been observed in our group, though in very different
experimental conditions, using polarized fluorescence detection
\cite{bou82}.  This earlier work provides valuable information: our detailed
analysis of the main systematic effects \cite{bou86}, transposed to the present situation, allows us to evaluate the
risks of systematic effects in this new detection scheme. As confirmed by other transverse field PV experiments, 
the most troublesome source of systematics arises from interference between the Stark-induced electric-dipole and the
magnetic-dipole amplitudes. Thanks to the use of two counterpropagating beams, there are two criteria to discriminate the $
E_1^{pv} $ against the
$M_1$ contribution, based on their different behaviour under
$\xi_{ex}$ and $\hat k_{ex}$ reversals. However this is not enough to guarantee the absence of systematic effects at 
the level needed for reaching a high accuracy. Indeed, if the cell input window has a birefringence $\alpha_{ex}^{(1)}$ with
its axes at $45^{\circ}$ to
$\vec E_{\perp}$, then the $6S-7S$ excitation rate is affected by a  $\xi_{ex}$-dependent contribution, the vapor acting like a
linear analyzer oriented along the direction of  
$\hat E_{\perp}$, which detects $   \vert  \hat E_{\perp} \cdot\hat
\epsilon_{ex}\vert^2 =   (\frac{1}{2} - \alpha^{(1)}_{ex} \, \xi_{ex} \,)$. To first approximation this contribution is the same
for the
$7S$ population and the $7S$ polarization signals, respectively proportional to $\cal{A}$ and $D_{at}$, so that it cancels in
their ratio. However, as soon as the contributions to the transition rate proportional to
$\alpha \beta $ and $\beta^2$ are taken into account this compensation is no longer exact and in the ratio 
$D_{at}/\cal{A}$,  the birefringence problem is not eliminated, but reduced at the level of 
$2 \alpha^{(1)}_{ex} \kappa$ (where
$\kappa= -\frac{\beta}{4 \alpha}+ \frac{5 \beta^2}{6 \alpha ^2}\simeq 3\times 10^{-2} $). With $2 \alpha^{(1)}_{ex} =
10^{-3}$ rad, the systematic effect would be $3\times 10^{-5} \times P^{(1)}  $, or 60$\%$ of the PV signal itself. 
Clearly other means of discrimination are necessary.  

\subsection{Recovering the axial symmetry}
A further efficient way to suppress this source of systematics would be to exchange the direction of the fast and 
slow axes of birefringence with respect to the direction of $\vec E_{\perp}$. Since this looks unfeasible with a cesium vapor
cell, we have considered instead the possibility of rotating the field by 90$^{\circ}$, which is of course equivalent. We
propose a new design of the electrodes (Fig. 3) such that by simply switching voltages and currents the measurement can be
repeated with both $\vec E_t$ and $\vec B$ fields rotated by
90$^{\circ}$. This is equivalent to switching the sign of $\alpha_{ex}^{(1)}$, so that the $\xi_{ex}$-dependent contribution to
$P^{(1)}$ due to
$\alpha_{ex}^{(1)}$ also changes sign.  
Owing to this additional reversal we expect to improve the
discrimination between
$P^{pv}$ and $P^{(1)}$ by more than two orders of magnitude. 
As shown in Fig. 3, it is even possible to perform rotations of $\vec E_{\perp}$ by successive increments of
$45^{\circ}$. First, this provides a sensitivity to the birefringence $\alpha^{(3)}_{ex} $ having its axes parallel and orthogonal to
$\vec E_{\perp}$ via the observation of the $\xi_{ex}$ dependence of $\cal{A}$ once $\vec E_t$ is rotated by 45$^{\circ}$.
Second, since
$\alpha^{(3)}_{ex}$ can combine with a small misalignment of $\vec E_{\perp}$ and $\vec B$ to generate a systematic effect, the
possibility of alternating measurements in two orthogonal directions of $\vec E_{ \perp}$ are here again very helpful. 

\vspace{3mm}
\begin{figure}
\centerline{\epsfxsize=50mm  \epsfbox{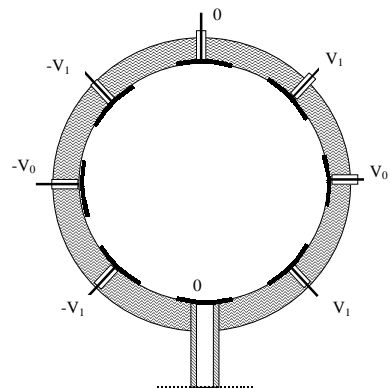}}
\vspace{5mm}
\caption{\small Electrode arrangement inside a cylindrical cell allowing one to rotate the transverse electric field by increments
of $45^{\circ}$. We have shown by numerical simulation that, by adjusting the potential ratio
$V_1/V_0$, we could make the field very homogeneous close to
the axis. The side arm is to be connected to the Cs reservoir (not shown). }
\end{figure}
The cell should be made from an insulating material, for instance high purity alumina to which sapphire windows 
can be glued \cite{sar89}. In our preliminary measurements, following ideas similar to those exploited for realizing the longitudinal
field configuration
\cite{jah01b}, we used an alumina cell with {\it external} electrodes. The design of Fig.3 with { \it internal electrodes} would provide
the possibility to eliminate the electric charges induced by photoionization of the sapphire windows under the
impact of the intense excitation pulse \cite{bou03}. However, it should be noted that, as confirmed by observation, the
transverse field configuration favours a  reduction of space charge effects. Indeed, due to its direction the field tends to drag
all the free electrons out of the interaction region. In addition if the cell length is increased, for a given excitation energy the
relative importance of the space charge effects located close to the windows will be reduced.

  At a first glance, the transverse field pump-probe experiment appears to lack cylindrical symmetry, one of
the main advantages inherent to the longitudinal $E$-field geometry. In actual fact, we can recover this symmetry by using
the electrode configuration proposed in Fig.3  which makes it possible to alternate measurements with directions of the
transverse $E$-field rotated by angular increments of 45$^{\circ}$ and directions of $\vec B$ simultaneously rotated by
adjusting the currents  in two pairs of coils. We mention that signals of atomic origin permit to test the defect of parallelism
between $\vec E$ and
$\vec B$ in each configuration \cite{noteC}.
\section{ Expected gain of sensitivity with $\vec E$ transverse}
\subsection{Estimated gain}  
 Now we have defined the operating conditions necessary for obtaining dark-field detection of a calibrated PV asymmetry
having a very specific signature in the transverse field configuration, we can discuss at last the crucial  objective at stake: how to
increase the experimental sensitivity by making use of the asymmetry amplification given by a 
longer interaction length. 
To estimate the SNR ratio for the polarimeter operating as a circular polarization analyzer we use the 
simplified model already employed to discuss asymmetry amplification in the $\vec E_{\parallel}$ configuration \cite{bou96}.
It provides clear analytical results for $dA_{LR}/d\theta$ and although we used it somewhat beyond its range of validity
(rather large  saturation levels, short probe pulse durations), we have found that its predictions reproduce our observations
well, not only  their general trend but even semi-quantitatively \cite{cha98}. (Note that numerical  methods to arrive at
quantitative predictions in specific conditions, if need be, have also been given in \cite{bou96}, but they lead to less transparent
results). Transposing to the transverse field experiment the reasoning leading to Eq. 1, we derive the expression of the Noise
Equivalent Polarization, $NEQP $, at the limit of quantum noise:
\begin{equation}
(NEQP)^{-1} = \sqrt{n_{in}} \exp{\left({\cal{A}}(E_{\perp})/2\right)} \times d A_{LR}/ dP. 
\end{equation} 
Since $P^{pv}$ is in the range of $10^{-6}$, the left-right asymmetry  $ A_{LR} \equiv D_{at}=
\frac{1}{2} (\exp{({\cal{A}} \zeta P}) - \exp{(-{\cal{A}} \zeta P}))$ can be written simply as $A_{LR} \simeq {\cal{A}} \zeta P$, 
which exhibits the sensitivity of $A_{LR}$ to measuring the $7S$  orientation.  $\zeta$ is an angular coefficient
depending on the probe transition, 
6/5 in the favorable case of the $7S_{F=4} \rightarrow 6P_{3/2, F=5} $ transition.       

 Writing explicitly the dependence of ${\cal{A}}(E_{\perp})$ on the adjustable parameters, ${\cal{A}}(E_{\perp}) = k
\alpha^2E_{\perp}^2 L_{\perp}$, such as the field magnitude and the cell length, we obtain the SNR/pulse in a
transverse field experiment:
\begin{eqnarray}
SNR(\vec E_{\perp}) &=&  \vert P^{pv}\vert \times (NEQP)^{-1} \,,\nonumber \\
&=& k {\rm Im} E_1^{pv} \alpha E_{\perp} L_{\perp} \exp{(\frac{k}{2}\alpha^2E_{\perp}^2 L_{\perp})} \times \sqrt{n_{in}}\,, 
\end{eqnarray} 
to be compared to the SNR/pulse in a longitudinal field case (where ${\cal{A}}(E_{\parallel}) \simeq  k
\beta^2 E_{\parallel}^2L_{\parallel}$, \cite{noteD}):
\begin{eqnarray}
SNR(\vec E_{\parallel}) &=& \vert \theta^{pv}\vert \times  (NEQA)^{-1} \,, \nonumber \\
 &=& k {\rm Im} E_1^{pv} \beta E_{\parallel} L_{\parallel} \exp{( \frac{k }{2} \beta^2 E_{\parallel}^2
L_{\parallel})}\times \sqrt{n_{in}}\,. 
\end{eqnarray}
The role of the interaction length is made clear from both Eqs. 5 and 6. Presently $L_{\parallel} = 8$ cm,  let us
suppose that we can operate with $L_{\perp} = 40$ cm and $E_{\perp} = \frac {\beta}{\alpha} E_{\parallel}$, we can then
reasonably expect:
\begin{equation}
SNR(E_{\perp}) =  \frac{40}{8} \exp{\left( \frac{{\cal{A}}(E_{\perp})}{2} -\frac {{\cal{A}}(E_{\parallel})}{2}\right )} \times
SNR(E_{\parallel}).
\end{equation}
With ${\cal{A}}(E_{\parallel}) \simeq 0.4 $ and  ${\cal{A}}(E_{\perp}) \simeq 2.0 $, Eq.7 predicts a sensitivity gain by a
factor of $\simeq 5 \times \exp{0.8} = $ 10 which, if real, would literally transform the experimental situation. 
One might wonder about using still larger values of  ${\cal{A}}(E_{\perp})$ by making 
$L$ longer or $E_{\perp} $ larger? In fact, at higher optical density, a new physical process comes into
play, namely superradiance.

 \subsection{Towards larger optical densities: spontaneous versus triggerred superradiance  }
At high optical densities ${\cal{A}}$, the excited vapor can superradiate
spontaneously within some time delay, which becomes shorter as ${\cal{A}}$ is increased. 
This process  has been the subject to both theoretical and experimental investigations in the past 
\cite{dic54,man95,har82}, and more recently for conditions corresponding to those encountered here \cite{bou91}.
 
The present situation is particularly well suited for studying superradiance: by smoothly increasing the optical
density, for instance with a knob driving the $E$-field magnitude, it is possible to study the atomic emission as it gradually
goes through three different regimes: i) linear amplification of a resonant saturating probe pulse, ii) triggered superradiance, when
superradiance emission is triggered by  the injection of a weak resonant probe pulse and iii) spontaneous superradiance occuring 
without any probe beam, within a delay $\tau_D$ .  In a first approximation, the delay of the onset of spontaneous superradiance
is inversely proportional to the number of excited atoms in the interaction region
$N$ and hence to $\cal{A}$: 
\begin{equation}
 \tau_D= \frac {8\pi}{3} \frac{a^2}{N \lambda^2 \Gamma} \log N\,. 
\end{equation}
Here $a$ is the diameter of the excitation beam,
$\lambda$ the wavelength of the probe transition and $\Gamma$ is the spontaneous emission rate 
for the most probable hyperfine component. 
Because of the near-absence of Doppler dephasing between the dipoles, due to the narrow spectral
width of the excitation laser, the threshold is nearly as low as Eq.8 predicts which is not the case  in many other practical
circumstances. However, we must bear in mind that the measurements reported in
\cite{bou91} were performed without any applied magnetic field. We expect such a field to cause dephasing and
hence delay the superradiant emission. Consequently, the values of
$\tau_D$ recorded versus the number of excited atoms, presented in \cite{bou91} only serve as a {\it lower limit} to
estimate the delay as a function of $\cal {A}$ in the conditions of this proposal.   

  \vspace{5mm}
\begin{figure}
\vspace{-7mm}
\centerline{\epsfxsize=45mm  \rotatebox{270}{\epsfbox{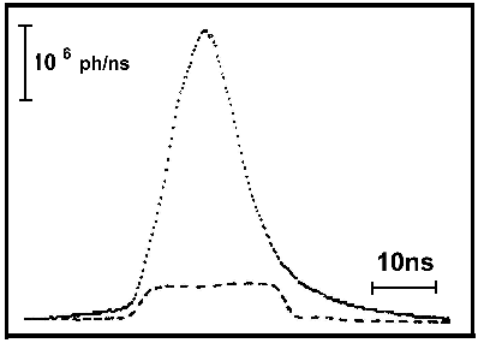}}}
\vspace{5mm}
\caption{\footnotesize Example of a triggered superradiant pulse. 
Dashed curve: the probe pulse (unamplified) in the absence of excited atoms. Dotted curve: due to the presence of excited
atoms the probe pulse is amplified. The experimental conditions ($n_{Cs}= 1.8
\times 10^{14} $ at/cm$^3$,  $E_{\perp} = 400 $ V/cm, $L_{\perp}= $25 mm) are such that the optical density is large. Note that a
non-zero intensity is observable even before the probe gate is opened. This is due to the photon leakage (a few $10^{-3}$)
through the ``closed'' probe gate, which prepares the build-up of the coherence of the atomic dipoles. When the probe gate is
opened (80
$\mu$W, 20 ns), full triggering of the superradiance occurs without any delay \protect \cite{bou91}.}   
\end{figure}

Actually, for PV measurements $\tau_D $ should not be too short (hence $\cal{A}$ should
not be too large) for two major reasons. First, to observe $\vec P^{pv}$, this vector must undergo a Larmor precession and   
$\tau_D$ must be long enough for the precession to have time to occur. Second, the random character of spontaneous
superradiance  deprives the detection of the stability required to make high precision measurements. Reliable 
measurements are possible only if the probe pulse can be applied before any superradiance spontaneously 
occurs. 

By contrast, the triggered superradiance regime offers very attractive possibilities: a very weak radiation
field is enough to suppress fluctuations in direction, frequency and polarization  of the spontaneous emission and to
give rise to complete, sudden emission of the whole 7S atom population leading to a {\it very large amplification
of the probe pulse} (see Fig.4). It is still more important to notice that {\it this is accompanied by an amplification of the left-right 
asymmetry} resulting from the angular anisotropy (orientation) of the radiating atomic medium \cite{bou91}. For instance, in a
situation close to that of Fig.4 we have observed an asymmetry amplification by a factor of 4. As underlined by Eq.5, 
such an increase in sensitivity of $d A_{LR}/ d P$ would be of invaluable help.   
 
After examination of our data giving $\tau_D$ versus ${\cal{A}}$, we arrive at the conclusion that the gain of
sensitivity by a factor of 10 predicted from Eq.7 with
$ {\cal{A}}(E_{\perp}) = 2.0 $, looks realistic, so that the future of high precision APV experiments based on amplification
by stimulated emission appear very promising. Using this technique measurements at the precision level of $0.1\%$ look
feasible.   
\section{An alternative experimental approach}
Finally, we would like to mention that the experimental configuration depicted in this paper presents
an interesting alternative. It can be verified that with an excitation beam {\it linearly polarized}, and with two {\it transverse}
electric and magnetic fields, which are {\it orthogonal}, then the excited atoms acquire an orientation in the
direction
$ \vec E\wedge \vec B$, hence along the direction of the excitation beam. For $\Delta F = 0 $ lines, this new
observable can be written as 
$\vec P^{pv}(B) \propto {\rm Im} E_1^{pv} \alpha \vec E \wedge \vec B $.  Contrary to the situation of section II, where we
considered an orientation
$\vec P^{pv}$ present in absence of  $B$, but requiring the $B$-field to be detectable by the probe, here we have a new  
orientation component directly created in the observation direction, but which cancels out when
$B=0$. To observe it one requires partial resolution of the Zeeman structure of the
$6S-7S$ transition which is not difficult in view of the Doppler width reduction obtained in the pump-probe
configuration (in a zero magnetic field the linewidth, partly due to the finite excitation duration, is about 80 MHz).  This has
several  advantages: the PV signal can be observed via  dark-field detection, using the differential two-channel polarimeter, {\it
without any pump-probe delay}; thus the risk of spontaneous superradiance is pushed towards higher optical densities. On
the $\Delta F$ = 0 transitions, with linear polarization $\hat \epsilon_{ex}$ chosen parallel to $\vec E$, 
there is no contribution arising from the $M_1$ amplitude. However to calibrate the $P^{pv}(B)$ signal, a good solution might
consist in switching the excitation polarization from linear to circular.  With a single excitation beam, the lifting of degeneracy
gives rise to a longitudinal polarization of the excited state $\vec P^{(1)}(B) \propto \alpha E M_1 \xi_{ex} \vec E \wedge \vec B$
due to the Stark-$M_1$ interference, which is detectable in the same conditions as
$\vec P^{pv}(B)$. The advantage is that, as in the first approach, this calibration gives access directly to ${\rm Im} E_1^{pv}/ M_1$
independently of the $E$-field magnitude and independently of the line shape. Note that, experimentally, there can be no confusion
with the transverse orientations $\vec P^{(1)}$ and $\vec P^{pv}$ discussed in sect. II, Eq.3, since their
Larmor precession is blocked by the
$\vec B$-field now parallel to the direction along which they are created.       

Moreover, let us point out that all that has been said in the preceding discussion about the
risks of systematic effects connected with window birefringence and the ways to eliminate them, can be transposed to
this configuration. In particular a rotation of the transverse fields around the common beam direction is
well adapted to overcoming the window birefringence. 

Due to its physical origin linked to partial resolution of the Zeeman structure, this orientation signal 
exhibits a line shape which is an odd function of the frequency detuning of the excitation beam with respect to the probe
beam frequency, and passes through zero close to resonance. The same result also holds if the signal observed is the circular
dichroism of the probe. In these conditions, one might prefer to observe the optical rotation signal which is an even function of the
frequency detuning
\cite{noteB}.
 
\section{Connections with previous work}
 In this section we would like to point out that the study of $E$-field assisted forbidden transitions and the search for
 atomic parity violation based on probe-beam detection, has opened up a wide class of magnetoelectric optical effects.
Up to now, several groups have been interested in the circular or linear dichroism which affect the beam exciting a
forbidden transition due to the contribution to the transition rate of an interference between the Stark-induced electric
dipole amplitude and either the
$E_1^{pv}$ or the
$M_1$ amplitudes. However, the forbidden transition is so weak that the 
dichroism is not directly detectable  on the {\it transmitted} excitation beam itself. In fact, the {\it population} of the excited
state has been the quantity monitored using  fluorescence detection. Thus, the circular dichroism
$\propto \xi_{ex} \hat k \cdot \vec E \wedge \vec B$ observed in the APV Cs experiment at Boulder \cite{woo97} was
detected by looking for a change of the fluorescence rate correlated with that of
$\xi_{ex}$. The linear dichroism $\propto (\vec E\cdot \hat \epsilon_{ex} \wedge\hat k)(\hat \epsilon_{ex}\cdot \vec
B)$ characteristic of a Stark-$M_1$ interference in the transition rate of a similar highly  forbidden Yb transition has also
been detected at Berkeley
\cite{bud03} via a change of the fluorescence rate correlated with the direction of the linear polarization of the excitation
beam with respect to the fields. The situation is different in  pump-probe experiments, {\it i.e.} our recent APV
experiment and those suggested here. The excited state presents some kind of angular anisotropy (alignment
\cite{bou96,bou03} or  orientation, see sect. II and IV)  
and {\it the magnetoelectric effect is revealed under the form of a linear or circular dichroism on the forward scattered
probe beam}. Probe-beam monitoring is well known in the context of 
optical pumping \cite{bud02}, but here in the context of highly forbidden transitions it is new to  
detect the rotational invariants which characterize the Stark interference terms appearing in the forbidden transition
probabilities under the form of magnetoelectric optical effects on the probe beam. The advantage is the amplification of these
effects  occuring while the probe beam  propagates through the excited medium with special enhancement when
collective effects take place.    
\section{Conclusion}
We have described a new strategy for performing APV measurements in a pump-probe experiment, 
using detection by stimulated emission following excitation {\it  in transverse electric and magnetic fields}.
This method takes advantage of experience gained from our present experiment using a
 longitudinal $E$-field, also based on stimulated-emission detection.
We have shown that one can circumvent the apparent breaking of rotational symmetry 
about the propagation axis by rotating the direction of the applied $E$ field.

By using a transverse $E$ field, one could increase both interaction length and field strength without the risk of discharge in a Cs
vapour. In this way, optical densities larger than those accessible in the longitudinal field geometry, can provide 
stronger benefit from the mechanism of asymmetry amplification during propagation of the probe beam in the excited medium. In
the limit of quantum noise, one can expect to gain in sensitivity by a factor of 10.  Even more favorable conditions can
be achieved: by choosing the
pump-probe delay to trigger superradiance of the anisotropic excited medium, the left-right asymmetry can be greatly amplified. 
With these intriguing possiblities, one could look forward to APV measurements at the
$0.1\%$ level.

\section*{References and notes}

$^*$ e-mail: Marie-Anne.Bouchiat@lkb.ens.fr

$^a${\small  Laboratoire Kastler Brossel is a Unit\'e de Recherche de l'Ecole Normale Sup\'erieure et de
l'Universit\'e Pierre et Marie Curie, associ\'ee au CNRS (UMR 8552). }

$^b${\small  F\'ed\'eration de Recherche du D\'epartement de Physique de  
l'Ecole Normale Sup\'erieure associ\'ee au CNRS (FR684)  }


\begin{thebibliography}{50}
\bibitem{noteF}
This, of course, does not hold any longer when the excited medium approaches the conditions of spontaneous 
superradiance, a situation where spontaneous emission causes additional noise, 
as discussed in sect. III C.
\bibitem{noteE}
 For matching the two beam intensities, a simple procedure might rely on a  highly parallel glass plate usable as a
temperature-tunable Fabry-Perot \cite{jah00}, close to normal incidence on the path of a single
beam. The range over which the transmission variation can be adjusted is wide (0 to 15$\%$) and with temperature
stabilization stable operating conditions can be achieved.
\bibitem{noteD}
The transverse and longitudinal field experiments that we consider here differ not only by the field direction,  
but also by the polarization of the excitation beam, circular versus linear and the hyperfine transitions involved, 
$6S_{F=4} \rightarrow 7S_{F=4} \rightarrow 6P_{3/2, F=5}$ versus $6S_{F=3} \rightarrow 7S_{F=4} 
\rightarrow 6P_{3/2, F=4}$. But it turns out that the overall change in the $k$ coefficient is small and can be neglected.
\end{thebibliography}
 \end{document}